\documentclass[pre,groupedaddress,showkeys,showpacs,twocolumn]{revtex4}
\usepackage{amsmath}
\usepackage{graphics}
\usepackage{graphicx}
\usepackage{amsfonts}
\usepackage{amssymb}
\usepackage{dcolumn}
\usepackage{bm}
\usepackage{natbib}
\usepackage{subfig}
\begin{document}
\title{Friction of rubber with surfaces patterned with rigid spherical asperities}
\author{D.T. Nguyen$^{1,2}$}
\author{S. Ramakrishna$^{2}$}
\author{C. Fretigny$^{1}$}
\author{N.D. Spencer$^{2}$}
\author{Y. Le Chenadec$^{3}$}
\author{A. Chateauminois$^{1}$}
\email[]{antoine.chateauminois@espci.fr}
\affiliation{1. Soft Matter Science and Engineering Laboratory (SIMM), UMR CNRS
7615,Ecole Sup\'erieure de Physique et Chimie Industrielles (ESPCI), Universit\'e Pierre et Marie Curie, Paris (UPMC), France\\
2. Laboratory for Surface Science and Technology, Department of Materials, ETH Zurich, Wolfgang-Pauli-Strasse 10, 8093 Zurich, Switzerland\\
3. Manufacture Francaise des Pneumatiques Michelin, Centre de Technologies, Clermont-Ferrand, France}
\begin{abstract}
This paper reports on the frictional properties of smooth rubber substrates sliding against rigid surfaces covered with various densities of colloidal nano-particles (average diameter 77 nm). Friction experiments were carried out using a transparent Poly(dimethyl siloxane) (PDMS) rubber contacting a silica lens with silica nano-particles sintered onto its surface. Using a previously described methodology (Nguyen \textit{et al.}, \textit{J. of Adhesion} \textbf{87} (2011) 235-250 ), surface shear stress and contact-pressure distribution within the contact were determined from a measurement of the displacement field at the surface of the PDMS elastomer. Addition of silica nano-particles results in a strong, pressure-independent enhancement of the frictional shear stress as compared to the smooth lens. The contribution of viscoelastic losses to these increased frictional properties is analyzed in the light of a numerical model that solves the contact problem between the rubber and the rough surface. An order-of-magnitude agreement is obtained between experimental and theoretical results, the latter showing that the calculation of viscoelastic dissipation within the contact is very sensitive to the details of the topography of the rigid asperities. 
\end{abstract}
\pacs{
     {46.50+d} {Tribology and Mechanical contacts}; 
     {62.20 Qp} {Friction, Tribology and Hardness}
}
\keywords{Friction, rough surfaces, Contact, Rubber, Elastomer, Torsion}
\maketitle
\section{Introduction}
\label{intro}
Rubber friction is a topic of huge practical importance in many applications, such as tires, rubber seals, conveyor belts and syringes, to mention only a few. As a consequence, frictional properties of soft elastomers have motivated several investigations for over half a century (for an historical perspective, the reader is referred to the paper by Sills \textit{et al.}~\cite{sills2007}). The velocity and temperature dependence of the frictional properties of commercial rubbers was first explored in early studies by mechanical engineers (see e.g. \cite{Ariano1930,roth1942,Thirion1946}). In a seminal work by Schallamach~\cite{schallamach1963}, this dependence was accounted for by thermally and stress-activated pinning/depinning mechanisms between rubber molecules and the contacting surface. While Schallamach focused on molecular processes at the frictional interface, other studies~\cite{bueche1959,greenwood1958} pointed out that a fraction of the energy dissipated during sliding motion is also due to viscoelastic losses resulting from the deformation of the soft rubber in the contact zone. These processes were first evidenced by Greenwood and Tabor~\cite{greenwood1958} in a series of experiments, in which hard spheres and cones were sliding or rolling on well-lubricated rubber surfaces. The selected lubrication conditions ensured that the thickness of the thin lubrication film was larger than the amplitude of surface roughness. Under such conditions, most of the friction force is assumed to arise from deformational losses within the rubber. These experimental results were analyzed using a simple model based on an empirical estimate of the fraction of the input elastic energy that is lost by hysteresis. This model was recently refined by Persson~\cite{persson2010}, using an approach in which the hysteretic losses are explicitly taken into account from the relaxation spectrum of the viscoelastic substrate.\\
Early experimental studies with single asperity contact were subsequently extended to the more complex situation of rubber sliding on microscopically rough surfaces. In a seminal work~\cite{Grosch1963a,grosch1963b}, Grosch examined the velocity and temperature dependence of the friction of filled rubbers against hard surfaces. In the case of rough tracks, a maximum in friction was found to occur at a sliding velocity related to the frequency with which the asperities of the rough surface deform the rubber surface. This maximum was absent on a smooth track, thus reflecting the deformation losses induced by the passage of the asperities over the rubber surface. From a theoretical point of view, Fourier methods of analysis can be employed to develop linear viscoelastic stress and displacement solutions for use in rough contact problems. Using such approaches, exact solutions for the deformation component of friction have been derived, as an example, for periodic arrays of identical asperities sliding against a power law viscoelastic materials~\cite{Schapery1978} or in the limiting case of a perfectly conforming contact between a rubber substrate and a stochastic surface~\cite{golden1980}. A more general contact-mechanics model for randomly rough surfaces was recently developed by Persson~\cite{Perrson2006,persson2001}. Using a spectral description of the topography of the rough surfaces, this theory predicts how the component of friction force associated with hysteretic losses varies with velocity and contact pressure from an estimate of the actual contact area. Some experimental results tend to support this theory~\cite{lorenz2011} but a detailed examination of the effects of surface topography on rubber friction remains very challenging in the case of randomly rough surfaces, whose characteristic length scales usually range over several order of magnitudes.\\
In this study, we take advantage of a technique developed by Huwiler~\textit{et al.}~\cite{Huwiler2007} and Kunzler~\textit{et al.}~\cite{kunzler2006}, which involves the sintering of colloidal silica nanoparticles onto silica surfaces. Using this technique, surfaces covered with various densities of spherical asperities with well-defined sizes and height distribution can be prepared. To some extent, such surfaces are reminiscent of the model surfaces considered in the rough contact theory by Greenwood and Williamson~\cite{greenwood1966}, in which spherical asperities with identical radius of curvature are assumed to be statistically distributed along the vertical direction. Experimentally, such patterned surfaces are of particular interest for rubber-friction studies because they offer the possibility to introduce roughness at a given length scale. Accordingly, the frequency distribution associated with the deformation of the rubber surface by the asperities is well controlled, as well as the volume of the viscoelastic substrate that is affected by hysteretic losses. This possibility is exploited here for a quantitative investigation of the hysteretic contributions to friction arising from localized viscoelastic dissipation at the nano-asperity scale. In a first section, the friction of such patterned silica surfaces against silicone rubber is investigated as a function of particle density, contact pressure and velocity. Using a previously developed contact-imaging methodology~\cite{nguyen2011}, the shear and pressure distributions at the frictional interface are determined from a measurement of the displacement field at the surface of the PDMS rubber. From these results, the pressure dependence of the frictional shear stress is discussed. In a second part, the experimental results are analyzed in the light of a theoretical contact model, which allows the role of viscoelastic losses associated with substrate deformation by the nano-asperities to be evaluated.
\section{Experimental and numerical details}
\label{exp_det}
\subsection{Materials}
\label{mat}
\begin{figure}
	\resizebox{0.5\textwidth}{!}{\includegraphics{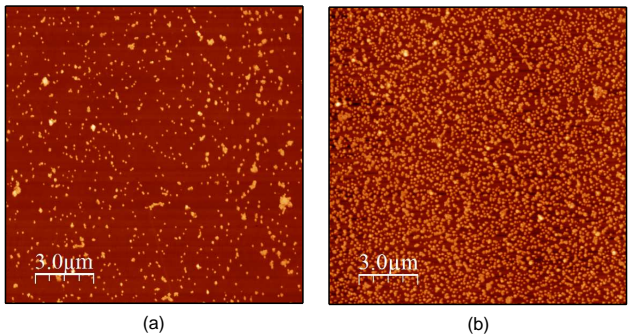}}
	\caption{AFM contact images of silica lenses decorated with sintered nano-particles. Adsorption time: (a) 10 min, (b) 30 min. The corresponding average particles densities are (a) $7.3$ particles/$\mu$m$^{2}$ and (b) 29.5 particles/$\mu$m$^{2}$.}
	\label{fig:AFM}
\end{figure}

A commercially available, transparent poly(dimethyl siloxane) (PDMS) silicone (Sylgard 184, Dow Corning, Midland, MI) was used as an elastomeric substrate. In order to monitor contact-induced surface displacements, a square network of small cylindrical holes (diameter 20 $\mu$m, depth 5 $\mu$m and center-to-center spacing 400 $\mu$m) was produced on the PDMS surface by means of conventional micro-lithography techniques (see reference~\cite{nguyen2011} for details). Under transmitted-light observation conditions, this pattern appears as a network of dark spots that are easily detected by means of image processing.  In order to prepare these marked PDMS surfaces, a resin template with a network of cylindrical pillars is first fabricated on a silicon wafer by means of soft micro lithography. The reactive silicone mixture in stoichiometric proportions (10:1 by weight) is then directly molded onto this template and cured in an oven at 70$^\circ$C for 48 hours. The specimen size is $6\:$cm$\times3\:$cm$\times1.5\:$cm. Before use, PDMS specimens were thoroughly washed with isopropanol and subsequently dried under vacuum.\\
Millimeter-sized contacts were achieved between the PDMS substrate and plano-convex silica lenses (Melles Griot, France) with a radius of curvature of 9.4~mm. The r.m.s. roughness of the as-received lens is about  0.3~nm, as measured from 1~$\times$~1~$\mu$m$^2$ AFM pictures. The silica lenses were decorated with sintered silica nano-particles using a previously developed procedure fully described in references \cite{Huwiler2007,kunzler2006}. The method relies upon the simple electrostatic attraction of negatively charged silica nanoparticles  onto the silica lens surface, previously rendered positively charged by coating with poly(ethylene imine). For that purpose, the coated lens was immersed in an aqueous silica nanoparticles suspension (diameter $\approx$~73~nm, purchased from Microspheres - Nanospheres, Cold Spring, NY). In order to achieve different particle densities, the immersion time of the lens was varied between 10 and 30 minutes. After particle adsorption, the lenses were dried with nitrogen and sintered at 1080$^\circ$C for two hours to remove any polymer on the surface and to partially sinter the nanoparticles to the surface. A smooth reference lens was also prepared using the same procedure, but without any nano-particles. This procedure ensured that the surface of the smooth lens was in the same physical and chemical state as that of the patterned surfaces. The patterned lenses were characterized by atomic force microscopy (AFM) and scanning electron microscopy (SEM). AFM images shown in Figure \ref{fig:AFM} indicate a uniform distribution of nanoparticles on the lens surface with only a few aggregates. The average density of the nanoparticles increases from 7.3 particles/$\mu$m$^{2}$ to 29.5 particles/$\mu$m$^{2}$ when the adsorption time is changed from 10 to 30 minutes. In AFM measurements, the tip end radius can be estimated to be of the order of magnitude of the nanoparticles sizes. As a result, it is not possible to extract any accurate information regarding the shape of the nanoparticles, except their heights relative to the lens surface. The measured average particle height above the substrate is 55~$\pm$~10 nm. The average particle diameter (77 nm) was obtained independently from SEM observations.
\subsection{Friction and displacement-field measurements}
\label{fric}
Friction experiments were carried out using a home-built device, which  is described in reference~\cite{Chateauminois2009}. Experiments were performed under imposed normal load (between 1.4 and 5.3 N) and velocity (between 0.01 and 1~mm~s$^{-1}$). The PDMS substrate was displaced with respect to the fixed glass lens by means of a linear translation stage. Lateral displacement and force were continuously monitored with a non-contact laser transducer (Keyence, France) and a strain gage transducer (Entran, France), respectively. Images of the contact zone were continuously recorded through the transparent PDMS substrate by means of a zoom lens and a CMOS camera. This system was configured to a frame size of 1024x1024 pixels with frame rates ranging from 0.3~Hz to 30~Hz. The measured friction forces were observed to vary slightly from one PDMS specimen to another and also as a function of the age of the specimen. As a consequence, all the experimental data to be compared in this paper were obtained using a single PDMS specimen and in a limited time.\\
The measurement of the surface lateral displacement field was based on the detection of the markers at the surface of the PDMS substrate by means of image processing. Image accumulation under steady-state sliding allows a spatial resolution of about 20~$\mu$m to be achieved. Vertical displacements within the contact zone were also deduced from a measurement of the indentation depth of the lens and a knowledge of its radius of curvature. The shear and contact-pressure distribution within the contact were determined from the measured displacement field with an already developed finite-element (FE) inversion procedure, taking into account the material and geometrical non-linearities of the problem. For full details regarding displacement field measurements and inversion procedure, the reader is sent to reference~\cite{nguyen2011}.
\section {Experimental results}
\label{exp_res}
\subsection{Friction \textit{vs} asperity density}
\label{asp_dens}
\begin{figure}
	\resizebox{0.45\textwidth}{!}{\includegraphics{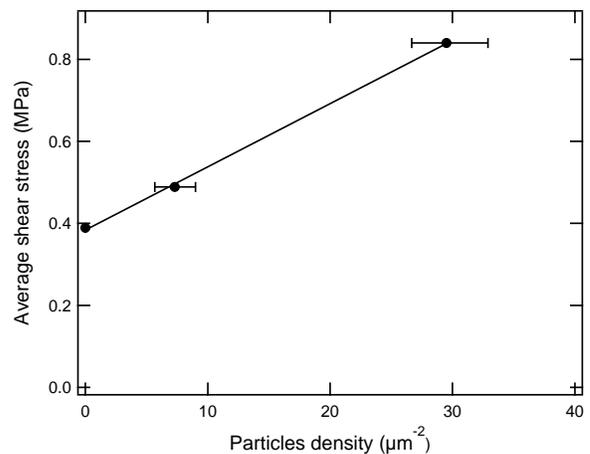}}
	\caption{Average frictional shear stress as a function of the nano-particle density (sliding velocity $v=1\: $mm s$^{-1}$, imposed normal force $F_N=$~1.4~N.}
	\label{fig:tau_density}
\end{figure}

\begin{figure}
	\resizebox{0.5\textwidth}{!}{\includegraphics{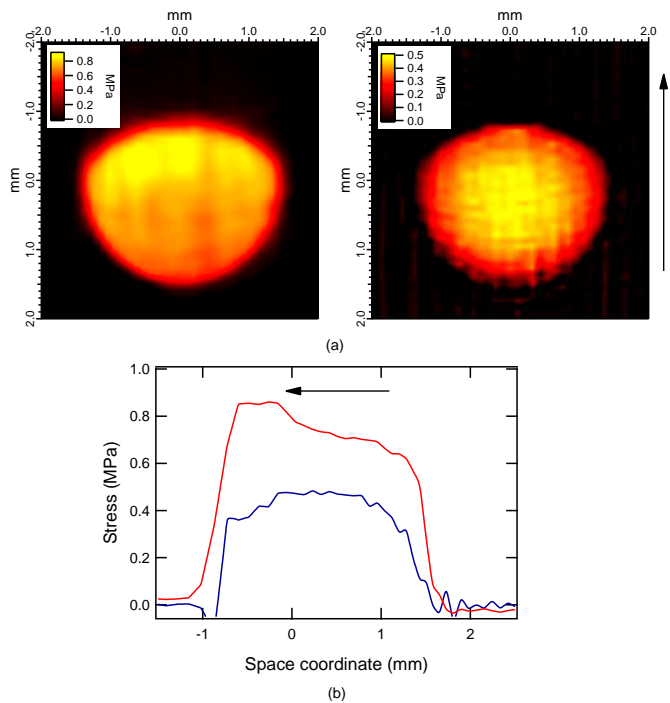}}
	\caption{Contact-stress distribution at the surface of a PDMS substrate sliding against a patterned silica lens (particle density $\phi=29.5$ particles/$\mu$m$^2$, $v=1\:$mm s$^{-1}$, $F_N=1.4\:$N). (a) shear stress (left) and contact pressure (right) fields deduced from the inversion of the measured displacements. The PDMS substrate is displaced from bottom to top with respect to the fixed glass lens as indicated by the arrow. (b) shear stress (red line) and contact pressure (blue line) profiles taken along the sliding direction and across the contact. The PDMS substrate is displaced from the right to the left as indicated by the arrow.}
	\label{fig:stress_field}
\end{figure}

\begin{figure}
	\resizebox{0.45\textwidth}{!}{\includegraphics{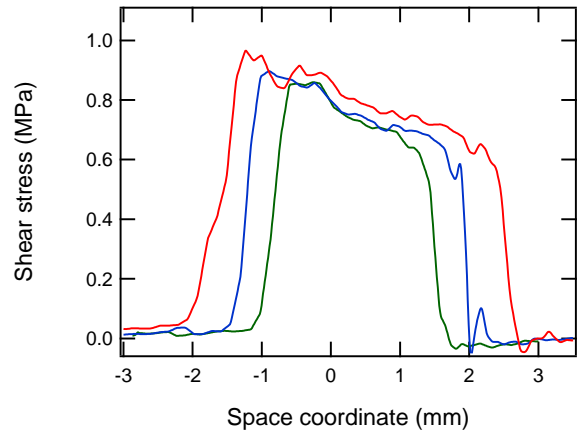}}
	\caption{Profiles of the shear stress  field along the sliding direction for various imposed normal loads, $F_N$ ($v=1\:$mm s$^{-1}$, particles density : $29.5$ particles/$\mu$m$^{2}$). Green: $F_N=1.4\:$N, blue: $F_N=3.4\:$N, red: $F_N=5.3\:$N. The PDMS substrate is displaced from the right to the left.}
	\label{fig:tau_pvar}
\end{figure}

Sintering nano-particles onto the silica lens systematically leads to an increase in the observed friction force. At the same time, the shape and area of the contact under steady-state sliding are also found to vary with the nano-particle density. In order to allow for a comparison between the various patterned surfaces, these changes in both the friction force and in the contact geometry were accounted for by considering the average frictional shear stress instead of the friction force. Here, the average shear stress is defined as $\overline{\tau}=F_T/A$, where $A$ is the measured macroscopîc contact area under steady-state sliding and $F_T$ is the friction force. Figure \ref{fig:tau_density} shows the change in the measured average frictional shear stress, $\overline{\tau}$, as a function of the nano-particle density under imposed normal load and sliding velocity. Patterning the silica surface results in a clear enhancement of the shear stress: increasing the particle density up to 30 particles/$\mu$m$^{2}$ results in a twofold increase in the frictional shear stress as compared to the smooth contact. Moreover, this increase with particle density is compatible with a linear relationship, suggesting that nano-particles contribute to friction independently of each other.
The distribution of the shear stress within the contact was further considered from the inversion of the measured surface-displacement field. Figure \ref{fig:stress_field} shows a typical example of the shear and contact-pressure distribution achieved under steady-state sliding. A maximum in the contact pressure is clearly visible at the center of the contact-reminiscent of a Hertzian distribution. As indicated by the profile in Figure \ref{fig:stress_field}b, the frictional shear stress exhibits a gradient along along the sliding direction which is uncorrelated to the pressure distribution. This gradient has already been been reported for similar contacts \cite{nguyen2011} and it will be the topic of a separate study. Within the framework of the present study, it is without significance as it does not alter the main conclusion that the shear-stress distribution within both smooth and rough contacts is independent of the contact pressure. This feature is preserved for both nano-particle densities.
As shown in Figure \ref{fig:tau_pvar}, this result is further confirmed by a series of experiments where the normal load is varied from 1.4 N to 5.3 N for a given particle density ($\phi=29.5$~particles/$\mu$m$^2$) without any change in the local shear stress within experimental accuracy. When two rough bodies are pressed together, contact is usually assumed to occur at discrete, localized, contact spots. Within such multi-contact interfaces, the level of local shear stress should therefore vary as a function of the contact pressure by virtue of the associated changes in the actual contact area. Here, the fact that the local shear stress does not vary with the contact pressure suggests that the contact between the smooth PDMS substrate and the patterned silica surface is nearly saturated, i.e. that the actual contact area does not significantly vary within the considered pressure range.
\subsection{Friction \textit{vs} velocity}
\label{velo}
\begin{figure}
	\resizebox{0.45\textwidth}{!}{\includegraphics{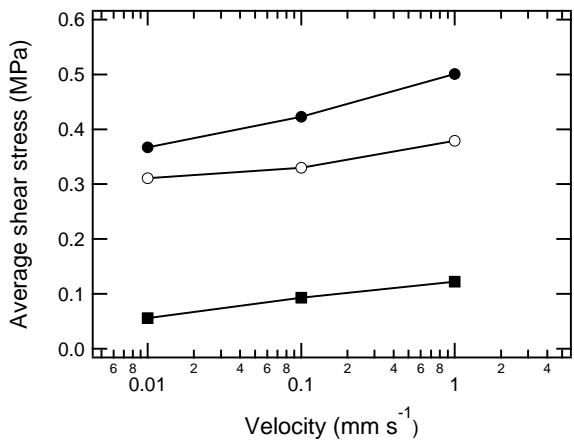}}
	\caption{Velocity dependence of the average shear stress of smooth and rough (particle density: 6.7~particles/$\mu$m$^{2}$) contact interfaces ($F_N=1.4$~N). ($\circ$) smooth lens, ($\bullet$) rough lens, ($\blacksquare$) difference between the average shear stress of the rough and smooth contacts.}
	\label{fig:tau_velo}
\end{figure}

Figure (\ref{fig:tau_velo}) shows the changes in the average contact shear stress as a function of the imposed sliding velocity for a smooth and a patterned ($\phi=6.7$~particles/$\mu$m$^{2}$) silica lens. In the case of the smooth lens, a weak, nearly logarithmic, velocity dependence is observed, as previously reported for similar smooth glass/PDMS contacts~\cite{chateauminois2010}. It turns out that the velocity dependence of the shear stress is slightly enhanced in the case of the patterned lens. This is further confirmed when the difference between the average shear stress of the rough and smooth contacts is considered, as shown by black squares in Figure \ref{fig:tau_velo}. A potential explanation for this effect would be that some additional viscoelastic dissipation is induced on the scale of the sliding nano-asperities, as a result of localized surface deformation of the PDMS rubber. Accordingly, the PDMS surface would be strained by the nano-particles at a characteristic frequency of the order of $v/d$ where $v$ is the sliding velocity and $d$ the particle diameter. Depending on the sliding velocity, strain frequencies in the range 10$^2$-10$^4$~Hz can be thus achieved locally on the PDMS surface in the sub-micrometer range. Although the selected PDMS is not a highly viscoelastic rubber, a significant increase in the shear loss modulus ($G"$) is measured by DMTA over such a frequency range (see Appendix A). Nano-asperity-scale viscoelastic contributions to friction could therefore-at least partially-account for the observed increase in the shear stress when nano-particles are present on the silica lens. This hypothesis is further considered in the following section.
\section {Rough-contact model}
\label{theo}
\begin{figure}
	\centering
	\subfloat[]{\label{(a)}\includegraphics[width=0.5\textwidth]{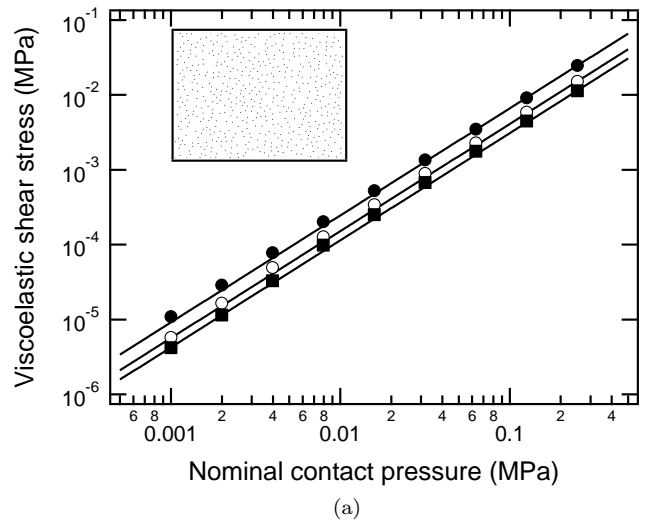}}\\
	\subfloat[]{\label{(b)}\includegraphics[width=0.5\textwidth]{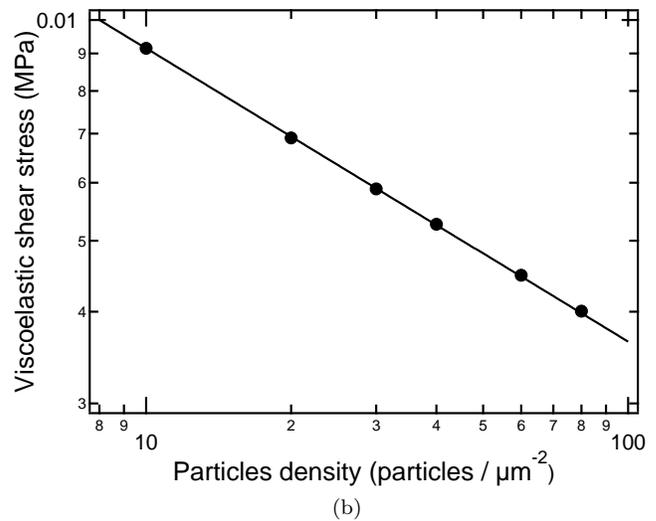}}
	\caption{Calculated viscoelastic shear stress for a contact in the suspended state ($v=1$~mm~s$^{-1}$).(a) shear stress as a function of contact pressure. ($\bullet$) $\phi=10\:\mu$m$^{-2}$; ($\circ$) $\phi=30\:\mu$m$^{-2}$; ($\blacksquare$) $\phi=60\:\mu$m$^{-2}$. Insert box: contact map ($3\:\mu$m~$\times 3\:\mu$m ) showing contact points as black dots. The black lines correspond to power-law fits with an exponent $1.43$. (b) shear stress as a function of particle density. The black line corresponds to a power-law fit with an exponent $-0.4$.}
	\label{fig:tau_visco_susp}
\end{figure}

The above results indicate that frictional shear stress is significantly enhanced by the presence of nano-particles on the surface of the silica lens. Moreover, this increase is compatible with a linear dependence on the particle density, suggesting that the contributions of nano-particles to friction are additive. In the following section, we will discuss these results within the framework of the classical Bowden and Tabor 'two term' model~\cite{Bowden1958,briscoe1981}. Accordingly, the frictional force is assumed to arise from two \textit{independent} contributions, denoted as the adhesive and the ploughing terms. The so-called adhesive term encompasses all the dissipative mechanisms occurring at the points of intimate contact between the solids, i.e. on length scales lower than the asperity size. The ploughing term corresponds to the force required to displace the rubber material from the front of the rigid nano-asperities. Here, it represents the contribution of the viscoelastic losses involved in the deformation of the rubber substrate by the nano-asperities. Rewritten in terms of shear stress, this model can be expressed as follows
\begin{equation}
	\tau=\tau_a+\tau_v
\end{equation}
where $\tau_a$ is the adhesive term and $\tau_v$ is the viscoelastic (ploughing) term corresponding to deformation at the nano-asperity scale. In a first approach, the viscoelastic component, $\tau_v$, can be evaluated by a simple scaling approach, in which the viscoelastic dissipation is assumed to occur within a volume of the order of $a^3$, where $a$ is the radius of the contact formed between the rubber substrate and a nano-asperity. The energy, $U$ dissipated during the deformation of a single asperity contact can thus be written as 
\begin{equation}
	U\approx E^{"} \epsilon^{2} a^3
\end{equation}
where $\epsilon$ is the average contact strain and $E^{"}$ is the loss component of the complex Young's viscoelastic modulus at some characteristic frequency of the order of $v/a$ where $v$ is the sliding velocity. Taking $\epsilon \approx a/R$ where $R$ is the radius of the nano-asperity it comes
\begin{equation}
	U \approx E^{"} \frac{a^{5}}{R^2}  
\end{equation}
The energy $U$ is dissipated when the asperity travels over a distance of the order of the contact size. The corresponding force can thus be written as
\begin{equation}
	f_v\approx\frac{dU}{da}=E^{"} \frac{a^{4}}{R^2} 
	\label{eq:force} 
\end{equation}
Within the assumption of non-interacting asperities, the total viscoelastic shear stress can thus be expressed as follows
\begin{equation}
	\tau_v \approx \phi E^{"} \frac{a^{4}}{R^2}
	\label{eq:pose}
\end{equation}
where $\phi$ is the number of asperities per unit surface area. Accordingly, the viscoelastic component of the frictional shear stress should be proportional to the particle density and to the loss modulus of the rubber substrate at the characteristic strain frequency imposed by the spherical asperities. The pressure dependence of the shear stress is embedded in the term $a^4/R^2$, which describes the local contact conditions on the asperity scale.\\
More refined simulations of the viscoelastic component of the frictional shear stress using the same ideas were carried out using a numerical model that solves the normal contact problem between a rigid rough surface and a smooth linear viscoelastic substrate.  In this approach, the friction force is assumed to arise only from the viscoelastic dissipation resulting from the deformation of the substrate under the action of the contact pressure. Then, the corresponding frictional shear stress can be derived explicitly as a function of the displacements in Fourier space. In addition, displacements can be related to the applied contact pressure from the expression of the Green's tensor~\cite{landau1986} in Fourier space~\cite{carbone2009,golden1980,Schapery1978} which simplifies to the following expression in the case of an incompressible substrate
\begin{equation}
	\tilde{u_z}=\tilde{G_z} \tilde{p}
\end{equation}
Here $\tilde{u}$ and $\tilde{p}$ denote the Fourier transforms of the vertical surface displacement and contact pressure, respectively. $\tilde{G_z}$ corresponds to the Fourier transform of the Green's tensor component along the vertical direction. The main issue remains the estimation of the displacement field, which is in general unknown unless an intimate contact is achieved between the surfaces. In order to determine the vertical displacement field, we used a numerical method to solve the viscoelastic normal-contact problem. The algorithm was initially proposed by Polonski and Keer~\cite{Polonski1999} and further detailed in reference~\cite{Allwood2005}. The conjugate gradient is used and the calculation of the displacement is operated in Fourier space. The calculation of the friction force is performed as described in reference~\cite{Schapery1978}. The problem is also written for steady-state conditions~\cite{johnson1985,shuangbiao2007} and periodic boundary conditions are introduced. The advantage of this method is to provide an exact spectral description of the deformed surface, from which viscoelastic dissipation can be estimated. In addition, this contact model is also able to handle potential effects arising from elastic coupling between neighboring asperities. In order to account for viscoelasticity, the elastic Young's modulus of the substrate is replaced by the complex viscoelastic modulus in the expression of the Green's tensor component. In a spirit similar to that of Persson's model \cite{persson2001}, this contact model is thus based on a spectral description of the surfaces, which can theoretically incorporate the entire frequency spectrum involved in the deformation of the viscoelastic substrate.\\
Two-dimensional calculations are carried out using a flat surface covered with a random distribution of identical asperities with hemispherical ($R=$~38.5~nm) caps and a height of 55~nm as determined experimentally. The viscoelastic properties of the PDMS rubber are described using a generalized Maxwell model whose parameters are fitted to the experimental data (see Appendix A). It can be noted in passing that, in the experiments, the selected specimen size ensures that semi-infinite contact conditions are achieved during sliding experiments (i.e. the ratio of the substrate thickness to the contact radius is greater than ten~\cite{gacoin2006}) in accordance with the theoretical contact model.\\
As a validation of the model, we first present some results corresponding to a suspended-state contact situation, in which contact only occurs at the top of the spherical capped asperities. From the results shown in Figure \ref{fig:tau_visco_susp}a, the relationship between the viscoelastic shear stress and contact pressure is seen to obey a power law dependence. The same conclusion holds for the dependence of the shear stress on the particles density (Figure \ref{fig:tau_visco_susp}b). Power-law fits of these numerical data provide $\tau_v \propto p^{1.43} \phi^{-0.4}$. These results can be compared to the theoretical prediction of equation (\ref{eq:pose}). Indeed, if a Hertzian contact is assumed to occur between the deformable substrate and the rigid asperities, equation (\ref{eq:pose}) can be rewritten as
\begin{equation}
	\tau_v \approx E^{"}R^{-2/3} \left( \frac{p}{E'} \right)^{4/3} \phi^{-1/3}
\end{equation}
where $E'$ is the storage component of the viscoelastic modulus at the characteristic loading frequency. Accordingly, $\tau \propto p^{1.33} \phi^{-0.33}$, which is very close to the prediction of numerical simulations. In the following section, we will consider a more complex situation, in which contact occurs between the rubber substrate and both the asperities and the base plane of the rough surface. This situation is likely to correspond to the experimental results and will be compared to the theoretical predictions.
\section {Comparison to experimental data and discussion}
\label{discussion}
\begin{figure}
	\centering
	\resizebox{0.5\textwidth}{!}{\includegraphics{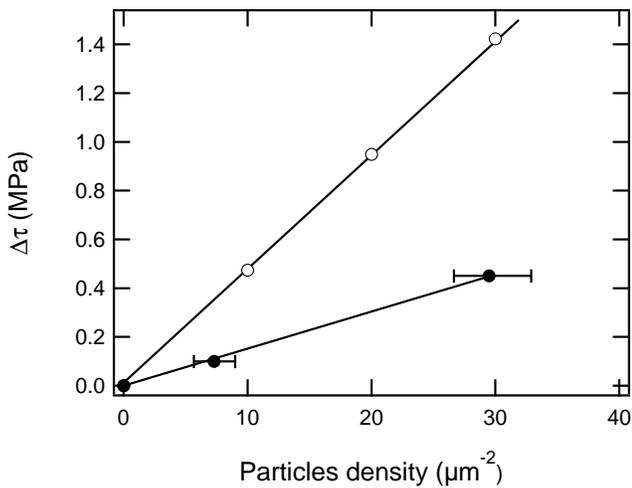}}
	\caption{Calculated and experimental viscoelastic shear stress as a function of particle density ($v=1$~mm~s$^{-1}$). ($\circ$) numerical simulation; ($\bullet$)  viscoelastic shear stress calculated as the difference between the measured shear stress of the patterned and smooth lenses.}
	\label{fig:delta_tau}
\end{figure}

\begin{figure}
	\resizebox{0.45\textwidth}{!}{\includegraphics{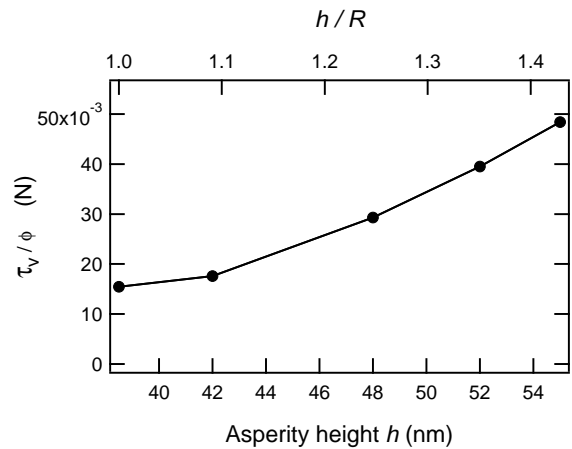}}
	\caption{Viscoelastic shear stress, $\tau_v$, normalized with respect to particle density, $\phi$, as a function of the height, $h$, of the spherical capped asperities (radius of curvature, $R$) above the reference plane ($v=1$~mm~s$^{-1}$).}
	\label{fig:asp_height}
\end{figure}

\begin{figure}
	\resizebox{0.45\textwidth}{!}{\includegraphics{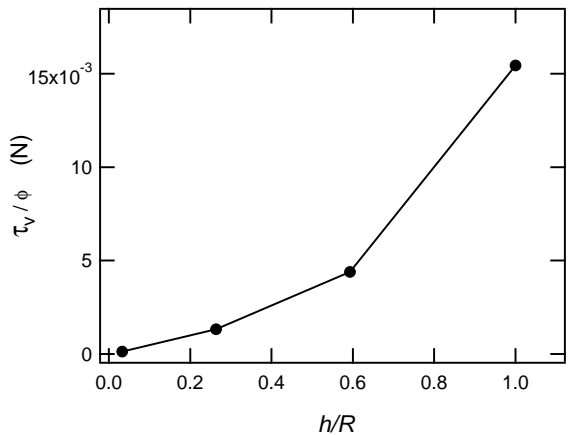}}
	\caption{Viscoelastic shear stress, $\tau_v$, normalized with respect to particle density, $\phi$, as a function of the aspect ratio, $h/R$ of the asperities where $h$ is the height above the reference plane and $R$ is the radius of curvature ($v=1$~mm~s$^{-1}$). The base diameter of the particles is kept constant and equal to 38.5~nm in all the simulations.}
	\label{fig:asp_shape}
\end{figure}

In order to extract the viscoelastic component, $\tau_v$, from the measured shear stress, we assimilate the adhesive component $\tau_a$ into the frictional shear stress measured with the smooth lens. This assumption implies that the enhancement of $\tau_a$ that arises from the increased area of intimate contact in the presence of nano-asperities is neglected. This hypothesis is justified by a simple calculation, which shows that for the highest asperity density (30~particles/$\mu$m$^2$), the maximum increase in the actual (intimate) contact area would be only 15\%, while the shear stress is increased by a factor of about two, compared to that on the smooth lens. Figure \ref{fig:delta_tau} shows the calculated and experimental viscoelastic shear stress as a function of particle density. For all these calculations, partial (unsaturated) contact conditions were found to occur between the rubber substrate and the rough surface with the rubber touching both the top of the asperities and some parts of the flat base plane. The linear relationship between $\tau_v$ and particle density is retrieved by the numerical simulations, but with a slope that is about 3 times higher than in the experimental case. However, this semi-quantitative agreement between experimental and simulated data is reasonable, if one considers all the uncertainties associated with the model parameters (such as particle shape and determination of the viscoelastic behaviour law).
In particular, it is interesting to consider the fluctuations in the calculated viscoelastic shear stress that are induced by changes in the height of the particles above the flat surface. Results reported in Figure \ref{fig:asp_height} show that the level of viscoelastic dissipation is very sensitive to this parameter: a 15~nm increase in the asperity height above the surface can result in a twofold increase in the viscoelastic shear stress.\\
It is also interesting to compare the numerical prediction of Figure \ref{fig:delta_tau} to a simple calculation using equation (\ref{eq:pose}) in the limiting case of an intimate contact between the rubber surface and hemispherical asperities. When $a \approx R$, this expression reduces to
\begin{equation}
	\tau_v \approx \phi G^{"} R^2
\end{equation}
Accordingly, the slope of the $\tau(\phi)$ relationship can be calculated from the radius of curvature of the nano-asperities and from the measured value of $G^{"}$ at the characteristic frequency defined by $v/R$. For $v=1$~mm~s$^{-1}$ and $R=$38.5~nm, the obtained value ($d\tau/d\phi \approx$~4~10$^{-9}$~N) is about one order of magnitude lower than the value of the numerical simulations ($d\tau/d\phi \approx$~5~10$^{-8}$~N). This discrepancy between the two calculations puts in question the relevance of the average frequency, $v/a$, to describe the viscoelastic response of the substrate at asperity scale. Depending on the shape of the asperity and on the contact condition, the strain frequency can in fact be distributed over a wide spectrum. This point becomes evident if the limiting case of hemispherical caps in intimate contact with the viscoelastic substrate is considered. In such a situation, infinite strain frequencies will be achieved at the periphery of the contact, while the frequency will vanish at the center of the contact. Such effects are also evidenced in more realistic numerical simulations, in which the ratio of the height, $h$, of the asperity to their radius of curvature, $R$, is varied at a constant asperity diameter (Figure \ref{fig:asp_shape}). A strong increase in the calculated viscoelastic shear stress is observed when $h/R \rightarrow 1$, i.e. when the tangent to the surface of the asperity becomes close to a vertical at the periphery of the asperity contact.
\section {Conclusion}
\label{concl}
In this paper, we have investigated the frictional properties of a smooth rubber substrate sliding on a rigid surface covered with mono-disperse colloidal asperities. Such 'model' rough surfaces with well-controlled asperity shape and size offer the possibility to revisit, in simplified contact situations, the current theoretical description of rubber friction with rough surfaces. Here, the emphasis was put on the so-called hysteretic component to friction that arises from the localized viscoelastic deformation of the rubber surface by the rigid asperities. We have found that the observed increase in the shear stress in the presence of colloidal asperities can be accounted for semi-quantitatively by a viscoelastic contact model that is based on a spectral description of the rough surfaces. However, it turns out that the calculated shear stress is highly sensitive to the geometrical details of the rigid asperities. In particular, the high-frequency strain components corresponding to elevated asperity slopes seem to make a dominant contribution to hysteretic friction. As consequence of the uncertainties regarding the actual asperity shape and height distribution, as well as the viscoelastic properties of the rubber, it seems unrealistic to expect better than an order-of-magnitude estimate of the shear stress, even for such simplified model surfaces. More generally, these results highlight the problem of the accuracy of the current theoretical predictions of hysteretic friction in the much more complex case of statistically rough surfaces. It is likely that the associated spectral description of the surfaces does not allow for the level of accuracy required to yield more than order-of-magnitude estimates of the hysteretic friction force. In addition, our contact model, as others, is based on a linear viscoelastic description of the rubber behavior. The contribution of the finite strains that are likely to be achieved within the contact remains to be evaluated.     
\begin{acknowledgements}
	Part of this work was supported by the National Research Agency (ANR) within the framework of the DYNALO project (project NT09 499845). SR and NDS wish to thank the ETH Research Commission for their financial assistance. The authors are also grateful to B. Bresson (SIMM) for his kind help with the AFM measurements.
\end{acknowledgements}
\appendix
\section{Viscoelastic properties of PDMS}
\begin{figure}
	\resizebox{0.45\textwidth}{!}{\includegraphics{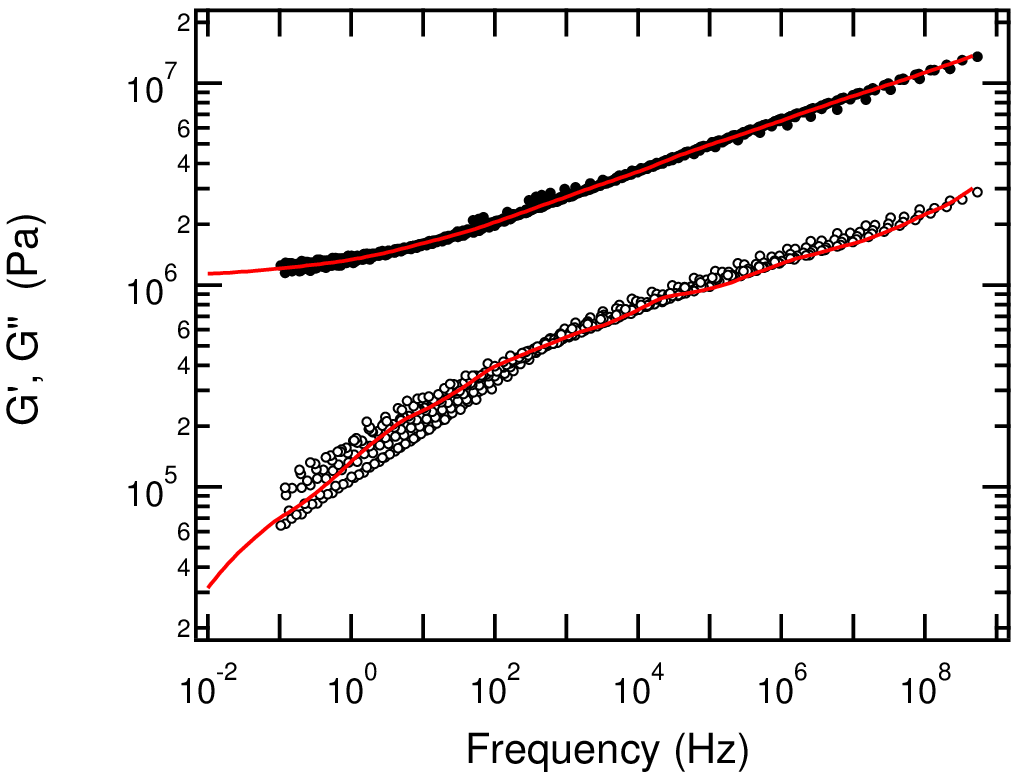}}
	\caption{Master curve giving the viscoelastic shear modulus of the PDMS rubber at 21$^\circ$C. Filled symbols: storage modulus $G'$; open symbols: loss modulus $G''$. The red solid lines correspond to the fit of the experimental data to a generalized Maxwell model}
	\label{fig:master_curve}
\end{figure}

The linear viscoelastic properties of the PDMS rubber were determined using Dynamical Mechanical Thermal Analysis (DMTA). PDMS disks (2 mm in thickness and 8 mm in diameter) are sheared at low strain (between 0.02\% and 0.05\% depending on the temperature) between the parallel plates of a rheometer (Anton Paar, MCR 501). Isothermal steps with 3$^\circ$C increments have been carried out between -77$^\circ$C and 23$^\circ$C. At each isothermal step, the shear modulus is measured during a frequency sweep between 0.01~Hz and 50~Hz after thermal equilibration of the specimen during 10~minutes. Figure \ref{fig:master_curve} shows the resulting master curve at a reference temperature of 21$^\circ$C. The solid lines correspond to the generalized Maxwell model fitted to the experimental data.
\bibliographystyle{unsrt}

%
%
\end{document}